\documentclass[journal]{IEEEtran}
\usepackage{cite}

\usepackage{amsmath, amsfonts}
\usepackage{graphicx}
\graphicspath{{./Images/}}
\usepackage{algorithm}
\usepackage{tikz}
\usepackage{algpseudocode}
\usetikzlibrary{arrows.meta,positioning,shapes.geometric,shapes.misc,fit,backgrounds}
\definecolor{inblue}{HTML}{2563EB}
\definecolor{capviolet}{HTML}{7C3AED} 
\definecolor{socyellow}{HTML}{CA8A04}

\definecolor{lossred}{HTML}{DC2626}
\definecolor{wire}{HTML}{111827}
 \usepackage{multirow}
\usepackage{nomencl}
\usepackage{hyperref}
\hypersetup{ colorlinks=true,linkcolor=blue, citecolor=blue, urlcolor=blue}
\usepackage{comment}
\usepackage{threeparttable}
\usepackage{orcidlink}

\newcommand{\soc}{\mathrm{SoC}}
\newcommand{\soh}{\mathrm{SoH}}
\newcommand{\uin}{\mathbf{u}}
\newcommand{\K}{\mathcal{K}}
\newcommand{\B}{\mathcal{B}}
\newcommand{\out}{\mathrm{out}}
\makenomenclature

\newcommand\copyrighttext{% 
\footnotesize This work has been submitted to the IEEE for possible publication. Copyright may be transferred without notice, after which this version may no longer be accessible.}
\newcommand\copyrightnotice{\begin{tikzpicture}[remember picture,overlay]
\node[anchor=south,yshift=10pt] at (current page.south) {\fbox{\parbox{\dimexpr\textwidth-\fboxsep-\fboxrule\relax}{\copyrighttext}}};
\end{tikzpicture}}

\begin{document}
\title{Operator-Theoretic Joint Estimation of Aging-Aware State of Charge and Control-Informed State of Health}

\author{ Rahmat K. Adesunkanmi\orcidlink{0000-0002-5483-5076}, Adel Alaeddini\orcidlink{0000-0003-4451-3150}, Mahesh Krishnamurthy~\IEEEmembership{Senior Member, IEEE}\orcidlink{0000-0002-1851-9666}\\ 
\thanks{Rahmat K. Adesunkanmi is with the Departments of Mechanical and Electrical Engineering, Southern Methodist University, Dallas, TX, 75205, USA (email: radesunkanmi@smu.edu)} 
\thanks{Adel Alaeddini is with the Department of Mechanical Engineering, Southern Methodist University, Dallas, TX, 75205, USA  (email: aalaeddini@smu.edu)}
\thanks{Mahesh Krishnamurthy is with the Department of Electrical Engineering, Southern Methodist University, Dallas, TX, 75205, USA}
\thanks{{\it (Corresponding author: Adel Alaeddini}.  This work was supported by U.S. National Science Foundation under grant ECCS-2538372 )}}

\maketitle
\copyrightnotice
\begin{abstract}
Accurate estimation of a battery’s state of charge and state of health is essential for safe and reliable battery management. Existing approaches often decouple these two states, lack stability guarantees, and exhibit limited generalization across operating conditions. This study introduces a unified operator-theoretic framework for aging-aware state of charge and control-informed state of health estimation. The architecture couples a Koopman-based latent dynamics model, which enables linear forecasting of nonlinear discharge-capacity evolution under varying operational conditions, with a neural operator that maps measurable intra-cycle signals to state of charge. The predicted discharge capacity is incorporated as a static correction within the neural operator pathway, yielding an age-aware state of charge estimate. Stability is ensured through spectral-radius clipping of the Koopman operator. The overall framework is trained end-to-end and evaluated on real-world lithium-ion battery datasets, demonstrating real-time capability while maintaining stable dynamics. To handle condition shifts and unseen regimes, the method integrates both zero-shot and few-shot out-of-distribution adaptation using only a limited number of cycles. Results show accurate and stable capacity forecasts, competitive state of charge trajectories on held-out cycles, and a direct, model-consistent mechanism for tracking capacity fade as a surrogate for state of health across diverse operating conditions.
\end{abstract}

 % \appendix
\nomenclature[R]{$t$}{Intra-cycle time}
\nomenclature[R]{$c$}{Cycle index}
\nomenclature[R]{$N_c$}{Number of intra-cycle samples in cycle $c$ ($t\in [1, N_c]$)}
\nomenclature[R]{$B$}{Number of batteries source in OOD training}
\nomenclature[R]{$Q,\, Q_n$}{Discharge / nominal capacity [$\mathrm{Ah}$]}
\nomenclature[R]{$Q^{\max}$}{Maximum discharge capacity [$\mathrm{Ah}$]}
\nomenclature[R]{$\soc_{c,t}$}{State of charge in $(c,t)$}
\nomenclature[R]{$\soh_c$}{State of health in cycle $c$}
\nomenclature[R]{$I$}{Current [$\mathrm{A}$]}
\nomenclature[R]{$V$}{Voltage [$\mathrm{V}$]}
\nomenclature[R]{$T$}{Temperature [$^\circ \mathrm{C}$]}
\nomenclature[R]{$\uin(\bar{\uin})$}{Intra-cycle (average) inputs $[V,\,I,\,T]$}
\nomenclature[R]{$x_{c,t}$}{Input vector at cycle $c$ and time $t$, combining $[V,I,T,\hat Q^{\max}_c]$}

\nomenclature[K]{$\varphi_k, \psi_k $}{Koopman lifting/ projection functions}
\nomenclature[K]{$z$}{Lifted latent state of $Q^{\max}$}
\nomenclature[K]{$\K$}{Koopman state transition operator}
\nomenclature[K]{$\B$}{Koopman input operator}
\nomenclature[K]{$\rho(\K)$}{Spectral radius of $\K$}
\nomenclature[K]{$r$}{Target stability radius for clipping}

\nomenclature[F]{$g_f$}{Fourier neural operator function}
\nomenclature[F]{$\varphi_f,\ \psi_f$}{FNO lifting / projection maps}
\nomenclature[F]{$v_\ell$}{FNO hidden field at layer $\ell$}
\nomenclature[F]{$\mathcal{F}_\ell$}{Fourier layer}
\nomenclature[F]{$R_\ell(k)$}{Complex FNO kernel at wavenumber $k$}
\nomenclature[F]{$W_\ell$}{Pointwise convolution for FNO}
\nomenclature[F]{$K_{\mathrm{modes}}$}{Number of Fourier modes}

\nomenclature[H]{$\theta$}{Trainable parameters ($\{\theta_{\varphi_k}, \theta_{\K}, \theta_{\B}, \theta_{\psi_k}, \theta_{g_f} \}$)}
\nomenclature[H]{$\alpha$}{Learning rate}
\nomenclature[D]{$\mathcal{D}_b$}{Dataset for multi-batteries $b$: $\{(\uin_{b,i},\,\out_{b,i})\}_{i=1}^{N_b}$}
\nomenclature[D]{$\mathcal{P}_b$}{Data distribution for battery $b$}

\printnomenclature

\section{Introduction and Related Work}
\IEEEPARstart{L}{ithium-ion} batteries (LiBs) are central to modern energy storage applications, from electric vehicles and portable electronic devices to stationary grid systems, due to their safety, operation lifespan, and energy density \cite{TAKENO2005298,SHEN2018161}. Over time, the internal reactions within the battery will change the maximum output current, resistance, and other parameters, leading to a fade in cycling stability and subsequently affecting battery life and performance \cite{qi2016lifetime}. Hence, to optimize performance and ensure safe operation, research on state of health (SoH), and State of Charge (SoC) estimation remains a priority for industrial deployment \cite{shu2021flexible,wang2024physics}. Joint estimation from operational data explicitly considers coupling between SoC and SoH because capacity fade alters SoC-voltage relationships and necessitates simultaneous updates of both states. Existing lines of work are primarily split into physics-based estimators grounded in equivalent-circuit (ECM) or fractional-order (FOM) models, utilizing Kalman filter (KF), as well as data-driven or hybrid models that fuse learned health indicators with model-based SoC tracking.

When the SoC is first scaled by the battery’s current cycle capacity, the open-circuit-voltage (OCV) curve stays almost the same for SoC values above the mid-range. This near-invariance is a key observability cue in prior work. It allows an ECM coupled with Kalman estimators to perform simultaneous SoC and capacity estimation across temperatures and dynamic profiles without requiring full cycles \cite{kadem2025co}. Building on the same observation, partial-interval OCV alignment reconstructs capacity and resets SoC by matching estimated OCV from two consecutive partial discharges to a temperature-adjusted reference \cite{9741297}. During deployment, real-time OCV identification in battery management systems with electrode-level reconstruction infers capacity and electrode aging from fragmented OCV under varying temperatures \cite{9818955}. Related approaches use OCV resets and recursive least-squares prediction for joint updates under dynamic loads and changing temperature \cite{en13071811,9195346}, and decouple OCV estimation from ECM parameters using dual adaptive $H_\infty$ filtering to improve robustness \cite{liu2018new}. These OCV-centric methods are attractive because they exploit identifiable events already present in operational data, such as relaxations, constant current-constant voltage tails, and near-cutoff segments, and because they can reduce the coupling between SoC scaling and capacity drift.

A second class of multiscale designs acknowledges that capacity evolves at a much slower timescale than SoC and that credible online estimators must separate the two dynamics while sharing information where necessary. Dual extended/unscented Kalman filters (EKF/UKF) or particle-filter estimators with ECM/FOM backbones update parameters online and have shown reliable joint performance under drive cycles and temperature variation \cite{9386077,er.8541,en15197416,9813961,2.12965}. Extensions address nonstationarity and noisy measurements via Sage-Husa and central-difference KFs, recursive total least squares for parameter identification, and Gaussian-sum particle filters to handle multimodality or heavy-tailed noise \cite{en17071640,10758433,2.12965}. To improve behavior and reduce bias under changing C-rates, fractional-order and rate-aware hierarchies that explicitly model polarization and diffusion effects are adopted \cite{9632433}.

Over the years, data-driven methods have emerged as powerful tools for simultaneously estimating SoC and capacity loss (or SOH), overcoming limitations of traditional physics-based approaches. These models utilize historical operational data to capture complex battery dynamics, eliminating the need for detailed knowledge of internal electrochemical processes. Hybrid models extract SoH-related indicators from voltage-current features using machine learning (such as convolutional neural networks (CNNs), hybrid kernel extreme learning machine (HKELM), or support vector machine (SVMs), and feed the inferred capacity to adaptive ECM+EKF/UKF for SoC estimation \cite{10586266,electronics14071290,27}. Long Short-Term Memory (LSTM) networks have also demonstrated the capability to capture temporal dependencies in concurrent and joint SoC/ SOH estimation \cite{lee2022estimation,li2023adaboost}. Recent advances improve upon standalone LSTM models by combining LSTM with CNNs for multi-feature fusion, where CNNs extract cycle features from battery data, and LSTMs learn aging characteristics over time \cite{fu2022state}. At the pack level, LSTM-based nonlinear state-space reconstruction has also been applied to mitigate cell imbalance and inconsistency, enabling more robust joint estimation across heterogeneous cells \cite{9}.

%Across studies, the literature generally indicates reliable joint tracking of SoC and SoH; nevertheless, models trained on specific battery chemistries and operating conditions often generalize poorly to new scenarios. Cell-to-cell variability and manufacturing inconsistencies further challenge model robustness.

Despite recent progress, prior joint estimation frameworks face fundamental challenges. OCV methods are easy to implement but are affected by incorrect initial SoC and sensor errors and require a long rest period to derive the OCV-SoC curve, limiting their suitability for real-time implementation\cite{coleman2008improved, xu2020improving}. Reliance on chemistry-specific OCV references or offline OCV fitting can limit generalization unless references are adapted online or reconstructed directly from telemetry \cite{9741297,en17133287,40}. The performance of physics-based approaches, including ECM and KF, depends heavily on the accuracy of model parameters and an understanding of internal battery dynamics\cite{s22031179}. In particular, the accuracy of KFs and their nonlinear variants depends on how well these simplified models represent the real cell and ECMs often require re-parameterization due to aging, temperature, and operating conditions \cite{knox2024advancing}.  In parallel, hybrid physics-based methods often remain specialized to specific chemistries or cycle regimes. Cross-chemistry and cross-format evidence for truly joint estimators is limited in comparison to capacity-only studies that exploit transfer learning or relaxation features \cite{36}. These limitations motivate the need for new frameworks that maintain stability, adapt to usage conditions, and support zero- and few-shot adaptation to unseen regimes.

Koopman and neural operators are emerging as means to combine stability and expressiveness. Koopman-based methods lift nonlinear dynamics into latent linear evolutions, enabling the application of linear control tools, with their applications have demonstrated that latent dynamics can disentangle the influence of aging and imposed control on battery trajectories \cite{garmaev2024deep,hossain2023data,10677261}. 

Complementing this, neural operators such as the Fourier neural operator (FNO) learn nonlinear mappings between function spaces by parameterizing integral kernels in the frequency domain. Unlike traditional convolutional or recurrent networks, FNO leverages Fast Fourier Transforms to efficiently encode long-range spatiotemporal dependencies \cite{doi:10.1137/21M1401243,fluids9110258}. This makes it particularly advantageous for battery segments like constant current–constant voltage (CC-CV) charging and post-charge relaxation, where voltage trajectories evolve gradually and nonlinearly over long durations. A key advantage is its resolution invariance, which enables SoC estimation without dependence on fixed time discretizations, making it robust to varying sampling and long rest intervals. Furthermore, a single trained FNO model generalizes well across temperatures and cell types, and can be adapted to new battery chemistries using transfer learning, an advantage not commonly found in standard deep learning-based SoC estimators\cite{xEV}. 

Together, these operator-theoretic advances emphasize the limitations of traditional physics-based methods in contemporary battery applications: dependence on fixed model structure and parameter calibration,  limited capacity to capture highly nonlinear dynamics, and an inability to adapt from large-scale operational data. Operator-theoretic models are promising alternatives for joint estimation because they offer a route to combine stability, interpretability, and data-driven expressiveness within a single end-to-end framework.

To this end, we propose an operator-theoretic framework that links SoC and SoH through an aging-aware capacity pathway, remains stable via a constrained latent operator, and is explicitly conditioned on control and operational parameters to support zero- and few-shot generalization. The framework couples a stable latent capacity forecast pathway, learned with a Koopman-inspired linear operator, with a neural-operator SoC pathway that processes measurable signals. Capacity forecasts are injected into the SoC pathway to render the SoC age-aware while supporting generalization and stability. Compared to OCV-centric and dual-filter pipelines, as well as hybrid ML-plus-filter systems, this approach maintains a clear backbone for stability, introducing explicit operational conditioning and lightweight adaptation mechanisms under domain shift. The main contributions are listed below. 

\subsection*{Our Contributions}
\begin{itemize}
\item \textbf{Aging-aware and control-informed joint estimator:} A unified, operator-theoretic architecture coupling a stable latent (Koopman-style) capacity pathway with a neural operator SoC pathway, making SoC explicitly age-aware and SoH control-informed.
\item \textbf{Operational conditioning with zero-/few-shot generalization:} Lightweight extrapolation to support zero-shot generalization and few-shot adaptable with minimal data.
\item \textbf{Stability guarantee:} Spectral-radius clipping on the latent operator and aging-aware correction of SoC.
\end{itemize}

The rest of this article is organized as follows: Section~\ref{Sec:Method} presents the mathematical framework of the proposed operator-theoretic learning framework; Section~\ref{Sec:Results} presents the simulation results and discusses the findings; and Section~\ref{Sec:Conclusions} summarizes the findings and concludes the paper.
\section{Operator-Theoretic Learning for Age-Aware Battery State Estimation}
\label{Sec:Method}
 Battery's SOH and SOC are related, so independently considering only one of these factors may result in estimation error. The prerequisite for accurate SOC estimation is the real-time adjustment of maximum available capacity. The SOH can be considered a constant during a single cycle and only gradually changes with long-term battery use, while the SOC exhibits noticeable changes during a single charge/discharge cycle, resulting in the two having different time scales\cite{11089775}. This section formalizes the joint estimation problem targeted by the proposed operator-theoretic framework. First, we restate SoC and SoH in an aging- and control-aware form, then describe how a Koopman-style latent dynamics model and an FNO are coupled so that the maximum discharge capacity (and thus SoH) forecasts inform the instantaneous SoC estimate. 

\subsection{Battery State Estimation Problem Definition}
A battery's SoC is the fraction of its extractable charge remaining at time $t$. With discharge capacity, $Q_t$, in Ampere-hour ($\mathrm{Ah}$), discharged at current $I_t$ in Amperes ($\mathrm{A}$), over an interval $[t_0,  t]$ and nominal capacity, $Q_n$ ($\mathrm{Ah}$), the SoC expression is defined as:  
\begin{align}
\soc_t
= \left(1 - \frac{Q_t}{Q_n} \right)\times 100\%, \label{eq:OGSoC}
\end{align}

\noindent where $Q_t= \int_{t_0}^t I_t dt$. Under aging, however, the usable capacity is reduced, so the correct normalization must depend on the current cycle ($c$) degradation level. Replacing $Q_n$ by the current maximum discharge capacity, $Q_c^{\max}$, yields an aging-aware SoC definition:
\begin{align}
\soc_{c,t}
= \left( 1 - \frac{Q_{c,t}}{Q_c^{\max}}\right)\times 100\%, \;  0 \le \soc_{c,t} \le 100.
\label{eq:SoC-aging-aware}
\end{align}

If $Q_c^{\max}$ is underestimated, SoC is biased high; if it is overestimated or kept at $Q_n$, SoC drifts downward between occasional OCV-based resets. Accurate online estimation of $Q_c^{\max}$ is therefore a prerequisite for stable SoC tracking under aging. 

Battery SoH indicates the overall condition and age of the battery and is defined in its capacity-based form as:
\begin{align}
\soh_c = \frac{Q_c^{\max}}{Q_n} \times 100\%, \;  0 \le \soh_c \le 100.
\label{eq:soh-def}
\end{align}

SoH is directly determined once $Q_c^{\max}$ is known, and this choice is consistent with battery management, where capacity fade is the most accessible proxy for overall health. $Q_c^{\max}$ is influenced by operating conditions such as temperature $T$, voltage $V$, and current $I$, and is also dependent on its previous state, $Q^{\max}_{c-1}$.

\subsection{Overview of the Proposed Framework for SoC and SoH Estimation}
Formally, we address the following \emph{joint SoC–SoH estimation problem}: given the current cycle's measured operational input parameters and maximum capacity, forecast the next cycle's maximum capacity, which incorporates the SoH information and a static age correction, and predict the aging-aware SoC under the next cycle's operating regime. For this method, we propose a two-level learning framework, by first employing a Koopman operator to learn the evolution of cycle-level maximum discharge capacity, which directly represents battery aging, before utilizing an FNO to estimate SoC at a temporal resolution within each cycle, while statically correcting for aging effects using the predicted capacity state. Mathematically, this problem is defined as finding the functions $f_k$ and $g_f$, the data-driven models that, respectively, embed cycle state into a linearized latent space to forecast the next state, and map intra-cycle measurements along with estimated discharge into an aging-aware SoC trajectory, optimizing the following equations: 
\begin{align}
\hat{Q}^{\max}_{c+1} &=f_k\big(Q_c^{\max}, \uin_c\big),\\
\hat{\soc}_{c+1, t} &= g_f\big((\uin_{c+1,t},\hat{Q}^{\max}_{c+1}\big)).
\end{align} 

\begin{figure*}[ht]
    \centering
    \includegraphics[width=\linewidth]{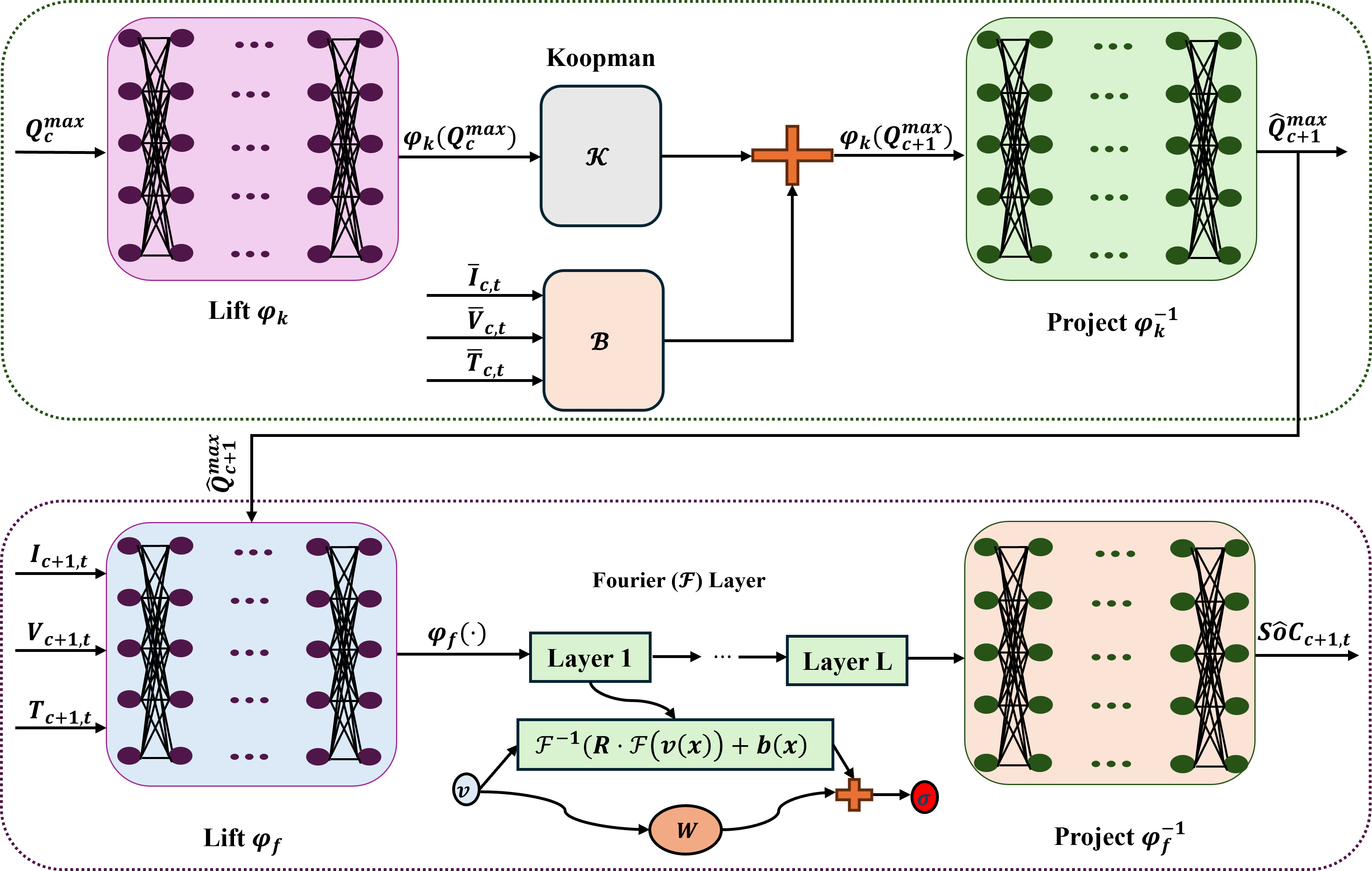}
    \caption{The proposed operator-theoretic joint estimation framework, composed of two coupled pathways: a Koopman-based latent dynamics model for aging-aware capacity forecasting (top) and an FNO-based SoC estimator (bottom)}
    \label{fig:KAEfno}
\end{figure*}

Here,  $\uin = \{V,I,T\}$, $f_k$ is a function dependent on the current state and the input, while $g_f$ is only reliant on input values. Fig.~\ref{fig:KAEfno} illustrates the proposed operator-theoretic joint estimation framework. The top pathway models the latent capacity dynamics using a Koopman-inspired architecture: the left block is a lifting neural encoder that maps inputs to a latent space; the central $\K$ and $\B$ modules represent the Koopman operator and control injection, respectively; and the right block is a projection decoder that maps the latent prediction back to capacity estimates. The bottom pathway is the SoC predictor based on an FNO: the lifting encoder maps raw input signals to a latent function space, the center module applies a Fourier layer, and the projection layer returns the final SoC estimate. 

\vspace{.03in}
\subsubsection{Cycle-Level Koopman Model for Maximum Capacity Forecasting} 
Consider a discrete-time nonlinear maximum capacity dynamics defined as
\begin{align}
  Q^{\max}_{c+1} = f_k(Q_c^{\max}, {\uin}_c), 
\end{align}

\noindent where $Q^{\max} \in R=\mathbb{R}^{n_k},{\uin} \in \mathbb{R}^{m_k \times N_c}$, and $f_k$ is the vector field. The Koopman operator assumes the existence of an infinite-dimensional space $N_k >> n_k$, real-valued nonlinear basis functions $\varphi_k$, which lift the original state space to the higher-dimensional state space so that the system in the lifted space follows the linear dynamics
\begin{align}\label{eq:kenc} 
 \varphi_k(Q^{\max}_{c+1}) &=\K\varphi_k(Q_c^{\max})+\B{\bar \uin}_c\\
\Leftrightarrow z_{c+1}& = \K z_c+\B {\bar \uin}_c,
 \end{align}

\noindent where $\K\in \mathbb{R}^{N_k\times N_k}$ and $\B\in \mathbb{R}^{N_k\times m_k}$ are the Koopman operators of the state and control spaces respectively\cite{bevanda2021Koopman, doi:10.1137/21M1401243}. For matrix compatibility, ${\bar \uin} \in \mathbb{R}^{m_k} $, the average of $\uin$, is used. The projection from the higher-dimensional space to the original lower-dimensional space is achieved by solving for $\psi_k =\varphi_k^{-1} \in \mathbb{R}^{n_k\times N_k}$ in the following equation:
\begin{align}\label{eq:kdec} 
\hat{Q}^{\max}_c =   \varphi_k^{-1}(z_c).
\end{align}

The process of discovering observable spaces has leveraged the capabilities of deep learning, replacing the traditional extended dynamic mode decomposition (EDMD)\cite{lusch2018deep}. Rather than guessing the right predefined dictionary of basis functions, a Koopman autoencoder is built, to learn $\varphi_k$, the encoding process (the observable function) that maps the state $Q_c^{\max}$ to $z_c$, the linear embedding operators $\K$ and $\B$, and $\psi_k$, the decoding function that projects the lifted space back to its original space all in an end-to-end manner.

\paragraph{\textbf{Encoder ($\varphi_k$) Construction}} The input $Q_c^{\max}$ is passed through a fully connected neural network (FCNN) with $N_k$ neurons and nonlinear activation function Scaled Exponential Linear Unit ($\mathrm{SELU}(x) = \lambda ( x \mathbf{1}_{\{x>0\}} + \alpha (e^x-1)\mathbf{1}_{\{x\le0\}} )$). This transformation lifts the capacity state to a higher-dimensional linear state.

\paragraph{\textbf{Linear Embedding}}
The lifted state $\varphi_k(Q_c^{\max})$ is passed through a linear embedding operation. The linear embedding consists of the Koopman operators $\K$ and $\B$, which are also trainable FCNNs but with no activation functions. $\K$ processes the lifted state, while the input  $\bar \uin$ is processed through $\B$. 

\paragraph{\textbf{Decoder ($\psi_k$) Construction}} To project to the
original state-space, a decoder $\psi_k$, an FCNN layer with $n_k$ neurons and SELU activation function is utilized. The decoder performs the inverse operation of the encoder, taking the higher-dimensional step-ahead state and mapping it to its original state. 

\vspace{.03in}
\subsubsection{Fourier Neural Operator for Age-Aware SoC Dynamics Estimation}
On top of this slow-timescale degradation model of Koopman, the intra-cycle SoC dynamics with an FNO, explicitly conditioned on the cycle's predicted $\hat{Q}^{\max}_{c+1}$, is estimated. Given paired input-output functions $(x,\soc)$ on a domain $D \subset \mathbb{R}^d$ (with $d=1$ in this study setting), the FNO learns an operator
\begin{align} 
g_f:\ x \mapsto \soc \quad\text{with}\quad \soc \approx g_f(x).
\end{align} 

FNO parameterizes ($g_f$) via a lifting function ($\varphi_f$), $L$ Fourier layers ($\{\mathcal{F}_\ell\}_{\ell=1}^L$) with nonlinearity ($\sigma$), and a projection function ($\psi_f$) such that:
\begin{align} 
v_0 &= \varphi_f(x)\\
v_{\ell+1} &= \sigma\big(\mathcal{F}_\ell v_\ell + W_\ell v_\ell \big), \quad  \ell=0,\dots,L-1,\\
\hat \soc &= \psi_f(v_L).
\end{align} 

Here ($v_\ell(\cdot):D\to\mathbb{R}^{c_\ell},\varphi_f,\psi_f,W_\ell$ are linear maps (1$\times$1 convolutions), while each $\mathcal{F}_\ell$ is a Fourier integral operator learned from data.

\paragraph{\textbf{Lifting ($\varphi_f$) Layer}}
At each intra-cycle time $t \in D$, an input feature vector is formed:
\begin{align} 
x_t = [\uin_{c+1, t}, \hat{Q}^{\max}_{c+1}], 
\end{align} 
\noindent of size $n_f$ (here $n_f = 4$), which is expanded into a higher-dimensional latent feature space of size $c_0$:
\begin{align}
\varphi_f(x_t) = W_{\varphi_f} x_t + b_{\varphi_f},  \quad W_{\varphi_f} \in \mathbb{R}^{c_0 \times n_f}.
\end{align} 

\paragraph{\textbf{Fourier ($\mathcal{F}_\ell$) Layer}}
For each channel, the Fourier transform in time is applied, multiplied by a learned mode-dependent kernel, truncated to the lowest $K$ modes, then inverse transform:
\begin{align} 
\mathcal{F}_\ell v_\ell = \mathcal{F}^{-1}\Big[ R_\ell(k) \cdot \mathcal{F}[v_\ell](k) \,\mathbf{1}_{\|k\|_\infty \le K} \Big],
\end{align} 
where $\mathcal{F}$ is the Fourier transform over the time domain, $R_\ell(k)\in \mathbb{C}^{c_{\ell+1}\times c_\ell}$ are learned complex weights, and the truncation $\|k\|_\infty \le K$ enforces spectral bias and efficiency. In this work, $\sigma$ is the Gaussian Error Linear Unit (GELU) activation function ($\sigma(x) = 0.5 x\cdot (1 + \operatorname{erf}(x/\sqrt{2})))$.

\paragraph{\textbf{Projection ($\psi_f$) Layer }}
The projection layer collapses the final hidden representation of size $c_L$ back to the output dimension $n_{\text{out}}$ ($1$ for SoC):
\begin{align}
\psi_f(v_L) = W_{\psi_f} v_L + b_{\psi_f},  \quad W_{\psi_f} \in \mathbb{R}^{n_{\text{out}} \times c_L}.
\end{align} 

\subsection{Loss Functions for End-to-End Training Objective}
In the Koopman encoder, to learn the invertible mapping from the state space to the observable space, the state reconstruction loss is defined using an $L_1 = \|\cdot\|_1$ (mean absolute error (MAE)) penalty:
\begin{align}
\mathcal{L}_\mathrm{Rec} = \|Q_c^{\max} - \psi_k \varphi_k({Q}_c^{\max}) \|_1, 
\end{align}

This loss trains the system state reconstruction capability at cycle $c$. To capture the linear dynamics in the observable space, the linear dynamic loss is defined as:
\begin{align}
\mathcal{L}_\mathrm{Lin}&= \left\|\varphi_k(Q^{\max}_{c+1})-(\K \varphi_k(Q_c^{\max}) + \B \bar{\uin}_c)\right\|_1.
\end{align}

Finally, the future state prediction loss is defined as: 
\begin{align}
\mathcal{L}_\mathrm{Pred} = \left\|Q^{\max}_{c+1}-\psi_k\left(\K \varphi_k(Q_c^{\max}) +  \B \bar{\uin}_c\right)\right\|_1. 
\end{align}

This loss promotes the training of the Koopman operator and the decoding dynamics by minimizing the prediction error of the next state. The FNO is optimized using a Huber loss between a true SoC value ($\soc_t$) and a predicted value ( $\hat{\soc}_t$) with the following piecewise function:
\begin{align}
\mathcal{L}_{\delta}&(\soc_t,\hat{\soc}_t) =\nonumber\\
&\begin{cases}
\tfrac{1}{2}(\soc_t-\hat{\soc}_t)^2 & |\soc_t-\hat{\soc}_t| \leq \delta \\
\delta \cdot \big(|\soc_t-\hat{\soc}_t| - \tfrac{1}{2}\delta\big) & |\soc_t-\hat{\soc}_t| > \delta
\end{cases}
\end{align}

\noindent where $\delta >$ 0 controls the transition between quadratic loss and linear loss. For a sequence of true and predicted SoC values over a cycle, the total loss is estimated to be 
\begin{align}
\mathcal{L}_{\soc} = \frac{1}{N_c} \sum_{t=1}^{N_c} \mathcal{L}_{\delta}.
\end{align}

Throughout the entire training process, the overall loss is defined, for weights of the loss components $\{\lambda_1, \lambda_2, \lambda_3, \lambda_4\} = \{1,1\times 10^{-4}, 1,1\}$, as $\mathcal{L} 
= \lambda_1 \mathcal{L}_\mathrm{Rec}
+ \lambda_2 \mathcal{L}_\mathrm{Lin} 
+ \lambda_3 \mathcal{L}_\mathrm{Pred} 
+ \lambda_4 \mathcal{L}_\soc$. $\lambda_1$ to $\lambda_4$ were selected via cross-validation on a held-out validation set. Empirically, the results remained stable over a broad range of these values. The training flow is illustrated in Fig.~\ref{fig:flow}.

\tikzset{
  font=\Large, % Paper-style font
  flow/.style    = {-{Latex[length=3mm,width=2mm]}, very thick, draw=wire},
  startstop/.style = {draw=black, fill=black!5, ellipse, minimum width=26mm, minimum height=10mm, very thick, align=center},
  process/.style = {draw=black, fill=#1!12, rounded corners=3pt, minimum width=80mm, minimum height=10mm, very thick, align=center},
  decision/.style= {diamond, draw=black, fill=gray!10, aspect=2, inner xsep=8pt, inner ysep=6pt, very thick, align=center},
  data/.style    = {draw=black, fill=#1!10, minimum width=80mm, minimum height=10mm, very thick, align=center},
  lossbox/.style = {draw=lossred!80!black, fill=lossred!10, rounded corners=3pt, minimum width=80mm, minimum height=10mm, very thick, align=center},
  % annot/.style   = {font=\scriptsize, text=black!80},
  group/.style   = {draw=black!30, rounded corners=6pt, inner sep=6pt},
}

\begin{figure}[!ht]
\centering
\resizebox{1\linewidth}{!}{
\begin{tikzpicture}[node distance=12mm and 24mm] 

\node[startstop] (start) {Start cycle $c$};

\node[process=inblue, below=10mm of start] (acq) {Acquire state and inputs\\
$x_c = Q^{\max}_c,\quad \uin_c=\{V_{c,t},I_{c,t},T_{c,t}\}_{t=1:N_c}$};

\node[process=capviolet, below=10mm of acq] (koop) {Stage I: Forecast head\\
Next state $Q^{\max}_{c+1} = f(Q^{\max}_c, {\bar \uin}_c)$};

\node[process=socyellow,  below=10mm of koop] (SoC) {Stage II: SoC head \\
$\hat{\soc}_{c+1, t} = g\big(\uin_{c+1,t}, \hat Q^{\max}_{c+1}\big)$};

\node[data=socyellow, below=10mm of SoC] (SoCout) {Output predicted values for cycle $c+1$\\ $\hat{Q}^{\max}_{c+1}, \, \hat{\soc}_{c+1}$};

% \node[process=inblue, left=10mm of SoC] (acq2) {Acquire inputs\\
% $\uin_{c+1}$};
\node[process=inblue, left=10mm of SoC, minimum width=25mm, minimum height=10mm] (acq2) {Acquire inputs\\
$\uin_{c+1}$};

\draw[flow] (start) -- (acq);
\draw[flow] (acq) -- (koop);
\draw[flow] (koop) -- (SoC);
\draw[flow] (acq2) -- (SoC);
\draw[flow] (SoC) -- (SoCout);

\node[lossbox, below=10mm of SoCout] (l_total) {Compute total loss\\
$\lambda_1\mathcal{L}_{\mathrm{Rec}}+\lambda_2\mathcal{L}_{\mathrm{Lin}}+\lambda_3\mathcal{L}_{\mathrm{Pred}} + \lambda_4 \mathcal{L}_{\soc}$};
\node[process=lossred, below=10mm of l_total, minimum width=80mm] (opt) {Backpropagation \& optimizer update};
% $\theta \leftarrow \theta - \alpha \nabla_\theta \mathcal{L}$};
\node[decision, below=10mm of opt, minimum width=24mm, minimum height=10mm] (early) {Early stop / Converged?};

\node[startstop, below=10mm of early] (end) {End};
\draw[flow] (SoCout)  -- (l_total);
\draw[flow] (l_total.south) -- (opt.north);
\draw[flow] (opt.south) -- (early.north);
\draw[flow] (early.south) --++(0mm,0) node[yshift=-5mm,  right] {Yes} -- (end.north);
\draw[flow] (early.west) -- ++(-49mm,0) node[xshift=10mm, above right] {No}  |- (koop.west);

% % ========= Background group boxes (drawn behind) =========
\begin{scope}[on background layer]
  \node[group, fit=(l_total)(opt)(early), 
        label={[xshift=10mm] above:\textbf{Training}}] {};
\end{scope}
\end{tikzpicture}}
    \caption{Flow chart of the training process of the methodology}
    \label{fig:flow}
\end{figure}
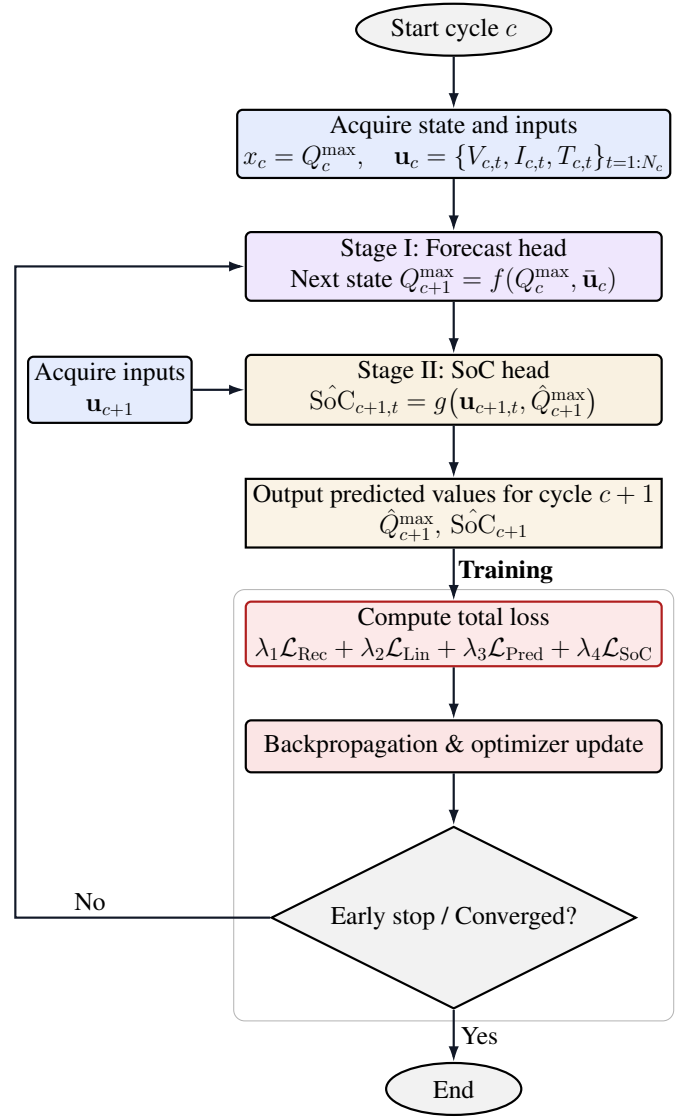

\subsection{Stability Analysis of the Learned Koopman Dynamics}
Stability checks are critical in Koopman-based learning because a linear operator $\K$ is fitted to approximate nonlinear dynamics in a latent space. The Koopman operator's latent states capture the battery's internal evolution in a lifted linear space, where each dimension (or eigenmode) reflects a distinct dynamical pattern. In the battery context, dominant eigenmodes often encode slow-varying degradation trends, such as capacity fade or impedance rise, while faster modes may reflect control-driven fluctuations or operational transients. The associated eigenvalues determine the temporal persistence of these effects: modes with eigenvalues close to one correspond to persistent degradation tendencies, whereas more transient effects decay rapidly. This decomposition not only stabilizes long-horizon forecasts but also provides insight into which degradation pathways dominate under different usage conditions. If $\K$ is unstable, predictions will diverge. For discrete-time systems, stability requires the eigenvalues of the learned Koopman operator to satisfy
\begin{align*}
  \rho(\K) = \max_i |\sigma_i(\K)| \leq 1. 
\end{align*}

The stability of the proposed architecture is analyzed by isolating the latent Koopman dynamics and treating the SoC head as a map driven by bounded inputs. Recall the lifted update in \eqref{eq:kenc} and \eqref{eq:kdec}. During training, the Koopman spectrum $\{\Lambda_i(\K)\}$ and the spectral radius $\rho(\K)$ are monitored; in deployment, the inputs are bounded via operating envelopes so that $\sup_c \|\bar{\uin}_c\|$ remains finite. Because the SoC head is strictly feedforward and receives a bounded scalar $\hat Q^{\max}_{c+1}$, it cannot destabilize the latent loop. To enforce these stability constraints in practice, spectral clipping is applied to the learned $\K$ after each gradient update. Let $\K = S D S^{-1}$ denote an eigen decomposition with $D = \operatorname{diag}(\Lambda_i)$. Any eigenvalue outside a prescribed radius $\rho_{\max} \leq 1$ is projected back onto the circle of radius $\rho_{\max}$:
\begin{align}
\tilde{\Lambda}_i = \begin{cases}
\Lambda_i, & |\Lambda_i| \le \rho_{\max},\\
\rho_{\max} \dfrac{\sigma_i}{|\Lambda_i|}, & |\Lambda_i| > \rho_{\max},
\end{cases} \end{align}

and reconstruct the clipped operator as $\tilde{\K} = S \tilde{D} S^{-1}$. This operation preserves the learned eigenvectors while guaranteeing $\rho(\tilde{\K}) \le \rho_{\max}$, thereby enforcing a stable linear dynamics model in the observable space.

\subsection{Optimization, Training and Evaluation Procedures}
The Koopman autoencoder and FNO components are optimized jointly using stochastic gradient-based methods. For the Koopman parameters, use the \texttt{Adam}optimizer. In contrast, the FNO parameters use the \texttt{AdamW} optimizer\cite{losh}, which decouples $L_2$ weight decay on the parameters from the gradient-based update. Let $\theta^{(n)}$ denote the (FNO) parameters at step $n$, $g^{(n)}=\nabla_\theta \mathcal{L}^{(n)}$ the mini-batch gradient, $\beta_1=0.9$ and $\beta_2=0.999$ the momentum coefficients. \texttt{AdamW} first forms exponential moving averages of the first and second moments,
\begin{align}
m^{(n)} &= \beta_1 m^{(n-1)} + (1-\beta_1) g^{(n)},\\
d^{(n)} &= \beta_2 d^{(n-1)} + (1-\beta_2)\, g^{(n)}\odot g^{(n)},
\end{align}
with bias corrections 
$\hat m^{(n)} = m^{(n)}/(1-\beta_1^n)$ and 
$\hat d^{(n)} = d^{(n)}/(1-\beta_2^n)$. 
\texttt{AdamW} then performs a decoupled weight-decay step followed by the adaptive update:
\begin{align}
\theta^{(n+\tfrac12)} &= \theta^{(n)} - \alpha \lambda \theta^{(n)},\\
\theta^{(n+1)} &= \theta^{(n+\tfrac12)} - \alpha \frac{\hat m^{(n)}}{\sqrt{\hat d^{(n)}}+\epsilon},
\end{align}
 
\noindent where $\alpha$ is the learning rate, $\lambda$ the weight-decay coefficient, and $\epsilon$ a small constant for numerical stability. Unless otherwise stated, $\epsilon=10^{-8}$ and weight decay $\lambda=1\times10^{-4}$, \texttt{StepLR} scheduler with a step size of $30$ and a decay factor $\gamma =0.5$, early stopping monitors validation loss with patience of $30$ and restores the best checkpoint. The Koopman parameters follow the same moment updates but without the decoupled weight-decay step, corresponding to the standard \texttt{Adam} optimizer. The proposed framework is evaluated using both the root mean square error (RMSE) and the MAE on maximum capacity forecasting and SoC trajectories. $Q_c^{\max}$ and $\hat Q_c^{\max}$ denote the true and predicted maximum capacity for cycle $c$, and let $N_{t}$ be the number of test cycles. The MAE and RMSE for the Koopman-based capacity forecast are defined, respectively, as:

\begin{equation}
\begin{aligned}
\frac{1}{N_t} \!\sum_{c=1}^{N_t} \big|Q_c^{\max} - \hat Q_c^{\max}\big|,  \quad
\sqrt{\frac{1}{N_t}\! \sum_{c=1}^{N_t} \big(Q_c^{\max} - \hat Q_c^{\max}\big)^2 }.
\end{aligned}
\end{equation}

For SoC prediction, $\soc_{c,t}$ and $\hat\soc_{c,t}$ are defined as the actual and predicted SoC at time step $t$ of cycle $c$ and $N_c$ the number of time steps in the cycle, then $N_{\text{tot}} = \sum_{c=1}^{N_t} N_c$ is the total number of evaluated time points. The MAE and RMSE respectively, are computed over all test trajectories as the ones for $Q_c^{\max}$, where $\frac{1}{N_t} \!\sum_{c=1}^{N_t} (\cdot)$ is replaced by $\frac{1}{N_{\text{tot}}} \sum_{c=1}^{N_t} \sum_{t=1}^{N_c}(\cdot)$. Algorithm~\ref{algm:1} summarizes the joint training procedure.
\begin{algorithm}[ht]
  \caption{Joint Training of Koopman-FNO Framework}\label{algm:1}
  \begin{algorithmic}[1]
    \Require Time-series inputs $\uin_c=[V_c,I_c,T_c]$ and maximum capacity $Q_c^{\max}$
    \Ensure Trained Koopman autoencoder and FNO  parameters $\{ \theta_{\varphi_k}, \theta_{\K}, \theta_{\B}, \theta_{\psi_k}, \theta_{g_f} \}$.

    \State \textbf{Train Koopman for capacity dynamics}
    \State Encode states: $z_c = \varphi_k(Q_c^{\max})$
    \State Evolve linearly: $z_{c+1}=\K z_c+\B\bar{\uin}_{c}$
    \State Decode: $\hat{Q}^{\max}_{c+1}=\psi_k(z_{c+1})$
    \State Concatenate predicted $\hat{Q}^{\max}_{c+1}$ with fast-time inputs:
      \[ x_t = [\uin_{c+1,t}, \hat{Q}^{\max}_{c+1}] \]

    \State \textbf{Train FNO for SoC dynamics}
    \State Apply FNO: $\hat{\soc}_t = g_f(x_t )$

    \State \textbf{Joint training objective}
    \State Combine Koopman and FNO losses:
\[\mathcal{L} = \lambda_1 \mathcal{L}_\mathrm{Rec} + \lambda_2 \mathcal{L}_\mathrm{Lin} + \lambda_3 \mathcal{L}_\mathrm{Pred} + \lambda_4 \mathcal{L}_\soc \]
    \State \textbf{Parameter update and spectral clipping}
    \State Update Koopman parameters with \texttt{Adam} and FNO parameters with \texttt{AdamW}
    \State Apply spectral clipping to $\K$ to enforce $\rho(\K) \le \rho_{\max}$
    \State \textbf{Inference}
    \State Predict capacity trajectory $\{\hat{Q}^{\max}_{c+1}\}$ via trained Koopman
    \State Compute SoC trajectory $\{\hat{\soc}_t\}$ via trained FNO
  \end{algorithmic}
\end{algorithm}

\subsection{Zero- \& Few-Shot  Out-of-distribution Generalization Across Batteries}\label{sec:method_gen}
Realistically, multiple batteries, possibly from the same chemistry and form factor but operated under slightly different profiles, are observed; hence, there is a need to create a pooled training objective. Let there be $B$ source batteries. For battery $b \in \{1,\dots,B\}$, let cycle-level input-output pairs be
\begin{align}
\mathcal{D}_b = \big\{ \big(\uin_{b,c}(\cdot),  \out_{b,c} \big) \big\}_{c=1}^{N_b},
\end{align}
where $\out_{b,c} =\{Q^{\max,b}_{c},\ \soc_{b,c}(\cdot)\}$ and $\soc_{b,c}(\cdot)$ denotes the SoC trajectory over the $N_c$ intra-cycle time steps for cycle $c$ of battery $b$. Assume each battery induces its own data distribution $\mathcal{P}_b$, so that $(\uin_{b,c}(\cdot), \out_{b,c}) \sim \mathcal{P}_b$, and these $\mathcal{P}_b$ are similar but not identical (i.e., different thermal environment and usage). A  single model with parameters $\theta = \{ \theta_{\varphi_k}, \theta_{\K}, \theta_{\B}, \theta_{\psi_k}, \theta_{g_f} \}$, is trained by minimizing the average loss over all observed batteries:
\begin{align}
\min_{\theta}  \frac{1}{B} \sum_{b=1}^B
\frac{1}{N_b} \sum_{i=1}^{N_b} \mathcal{L}_{b,c}
\label{eq:multi-batt-risk}
\end{align}

Equation~\eqref{eq:multi-batt-risk} is intended to inform the model of the capacity dynamics and SoC mapping that work reasonably well for all batteries observed so far. 

Suppose there is an unseen battery $b^\star$ with data distribution $\mathcal{P}_{b^\star}$ that was not present in training. In the zero-shot setting, the model is not adapted to $b^\star$ but simply evaluated as follows:
\begin{align}
\hat Q^{\max}_{b^\star, c+1} &= \psi_k \big( \mathcal{ K}\varphi_k(Q^{\max}_{b^\star,c}) + \B \uin_{b^\star,c} \big), \\
\hat{\soc}_{b^\star, c+1}(\cdot) &= g\big( \uin_{b^\star, c+1}(\cdot), \hat Q^{\max}_{b^\star,c+1} \big).
\end{align}
$\bar{\uin}_{b^\star,c}$ is the cycle-aggregated input and $\uin_{b^\star,c+1}(\cdot)$ denotes the intra-cycle input trajectory for the $(c+1)$-th cycle. Generalization here relies on the fact that \eqref{eq:multi-batt-risk} forced the backbone to see inter-battery variation during training, so $b^\star$ is just another sample from the family of batteries. 

A small adaptation step using only a handful of labeled cycles from $b^\star$ is also performed. Let
\begin{align}
\mathcal{S}_{b^\star} = \big\{ \big(\uin_{b^\star,c}(\cdot), \out_{b^\star,c}\big) \big\}_{c=1}^M
\end{align}
be the few-shot set, with $M \ll N_{b^\star}$. The pooled training dataset is updated to $\mathcal{D}' = \bigcup_{b=1}^B \mathcal{D}_b \ \cup\  \mathcal{S}_{b^\star}$, after which $\theta$ is further optimized.
\section{Implementation, Results and Discussions}\label{Sec:Results}
In this section, the proposed methodology is validated with the LiB dataset, beginning with an overview of the data and its preprocessing, followed by the presentation of the main results.

\subsection{Data Preprocessing}
A total of seven LiBs obtained from the research, published by \cite{jiangongzhu20226405084}, were employed to evaluate the proposed framework. The experimental dataset consists of cycling data for three 18650-type batteries with different chemistries (NCA, NMC, and a blended NMC+NCA). The cells were operated under a range of conditions in a temperature-controlled chamber, including multiple charging/discharging profiles and varying ambient temperatures. Supplementary material provided shows that all three batteries have a nominal voltage of $3.6\,\mathrm{V}$. For the NCA and NMC batteries, $1\,\mathrm{C}$ corresponds to $3.5\,\mathrm{A}$, whereas for the blended NMC+NCA cell, $1\,\mathrm{C}$ equals $2.5\,\mathrm{A}$. The cutoff voltage for the NCA batteries is $2.65V-4.2\,\mathrm{V}$ while that of the NMC and NMC+NCA is $2.5V-4.2\,\mathrm{V}$\cite{jiangongzhu20226405084}. The rest of the information for the selected dataset, after preprocessing, is summarized in Table~\ref{tab:data},

\begin{table}[h]
\centering
\renewcommand{\arraystretch}{1.1} \setlength{\tabcolsep}{3pt}
\caption{Technical specifications of Li-ion cells used in this study (see \cite{jiangongzhu20226405084})} \label{tab:data}
\resizebox{\linewidth}{!}{
\begin{tabular}{|l|c|c|c|c|c|c|c|c|} \hline
\multicolumn{1}{|c|}{Specification} & \textbf{B-1} & \textbf{B-2} & \textbf{B-3}& \textbf{B-4} &\textbf{B-5}&\textbf{B-6}& \textbf{B-7} \\ \hline\hline
Chemistry & NMC & NMC & NMC &NCA &NCA&NCA&NMC+NCA \\ \hline
Charge / Discharge Rate ($\mathrm{C}$) & $0.5 / 1$& $0.5 / 1$ & $0.5 / 1$ & $025/1$& $0.5/1$&$1/1$&0.5/1 \\ \hline
Temperature ($^\circ\mathrm{C}$) & $25$ & $35$ & $45$ & $25$& $25$&$25$&$25$ \\ \hline
Number of Cycles &$480$ &$1152$ & $200$&$339$ &$194$&$32$&$936$ \\
\hline
\end{tabular}}
\end{table}

Each raw cycle trace was resampled to a fixed temporal resolution to ensure comparability across cycles of different sampling rates. The original recorded measurements were mapped onto a uniformly spaced grid of length ($N_c$), spanning from the cycle start to the end. All selected features and target variables were linearly interpolated onto this uniform grid, preserving the overall shape of the dynamics. This produced, for every cycle, a fixed-size matrix ($\in \mathbb{R}^{N_c \times (\cdot)}$) of features and of SoC targets. All input features for both prediction outputs are normalized to ($[0,1]$) using \texttt{min-max} scaling: $X = \frac{X - X_{\min}}{X_{\max} - X_{\min}}$. The results are organized into two main parts. First, in Subsection~\ref{subsec:single}, the proposed approach is benchmarked against other established methods using a single battery case to provide a consistent baseline and examine the eigenvalue spectrum of the learned operator, assessing the stability of the underlying dynamics. Subsection~\ref{subsec:generalization} assesses how well the model generalizes across diverse operating conditions, broken down into three sub-parts: multi-temperature analysis, multi-C-rate experiments, and evaluation under different chemistries.

\subsection{Single Battery Benchmarking}\label{subsec:single}
To establish a reference case before exploring stability and generalization, the proposed method is compared to a range of existing machine learning baselines (excluding age for SoC) on a single battery (\textbf{B-1}). Traditional feedforward neural networks (FFN), recurrent neural networks (RNN), bidirectional LSTM (BiLSTM), CNN, and Temporal Convolutional Networks (TCN) are compared with the proposed coupled and decoupled framework. The traditional models are trained with the \texttt{Adam} optimizer and MSE as loss functions. Each model is tuned with appropriate hyperparameters, and its configurations, including batch size, learning rate, sequence length, network depth, and parameter count, are summarized in Table~\ref{tab:hyper}. 

\begin{table}[!ht]
\caption{Hyperparameters and parameter count of baseline and proposed models used in \textbf{B-1} benchmarking study.}\label{tab:hyper}
\centering
\renewcommand{\arraystretch}{1.1} \setlength{\tabcolsep}{3pt}

\resizebox{1\linewidth}{!}{
\begin{threeparttable}
\begin{tabular}{|c|l|c|c|c|l|c|} \hline 
Task & \multicolumn{1}{|c|}{Model} & Batch& Learning & Sequence & \multicolumn{1}{|c|}{Architecture} & Parameter\\ 

& & Size& Rate & Length&&Count\\ \hline \hline

\multirow{7}{*}{SoC} & FFN & $256$ & $5e-3$ &-& Dense: $256,128,64,1$; ReLU\tnote{1}&$42,241$\\ \cline{2-7} 

& RNN & $500$ & $5e-3$ & $10$ & RNN: $128,64$; Dense: $1$&$29,313$\\ \cline{2-7}

& BiLSTM & $500$ & $5e-3$ &$10$ & BiLSTM: $128,64$; Dense: $1$& $299,649$ \\ \cline{2-7}

&CNN &$500$& $5e-3$ & $10$& CNN1D: $256,128$; GAP\tnote{2} ; Kernel$=5$; & $176,897$ \\
&&&&& Dense: $64,1$; ReLU &\\ \cline{2-7}

& FNO& $6(4)$& $5e-4$ &$ N_c$&$ K=20$; $L=4$; Hidden$=48$; & $137,713$\\
&De(coupled)&&($5e-3$)&& Lift/Project$=32$; Dense: $1$ &\\ \hline

\multirow{9}{*}{$Q^{\max}$} & RNN & $2$ & $5e-4$ & $10$ & RNN: $32,16$; Dense: $1$ & $1,985$\\ \cline{2-7}

&CNN &$32$&  $5e-4$ & $10$& CNN1D: $32,16$; GAP; Kernel$=3$; & $2,113$ \\
&&&&&Dense: $8,1$; ReLU &\\ \cline{2-7}

& BiLSTM & $8$ & $5e-4$ &$10$ & BiLSTM: $256,128$; & $1,207,425$\\ 

&&&&& Dropout: $0.3$; Dense: $1$ & \\ \cline{2-7}

&TCN & $2$ & $5e-3$&$10$ & Dilation: $32,16,8$; kernel=3; & $64,448$ \\
&&&&& ReLU; Dense: $1$ &\\ \cline{2-7}

& Koopman & $3$ & $1e-4$& $1$& Encoder: $128, 64, 32, 16$;& $14,993$\\
&De(coupled)&&&&$ N_k = 32$; Decoder: $32, 16$ &\\ \hline

\end{tabular}

\begin{tablenotes}
\item[1] Rectified Linear Unit: $f(x) = max(0, x)$
\item[2] Global Average Pooling
\end{tablenotes}

\end{threeparttable}}
\end{table}

\begin{figure}[ht]
 \centering
 \includegraphics[width=\linewidth, ]{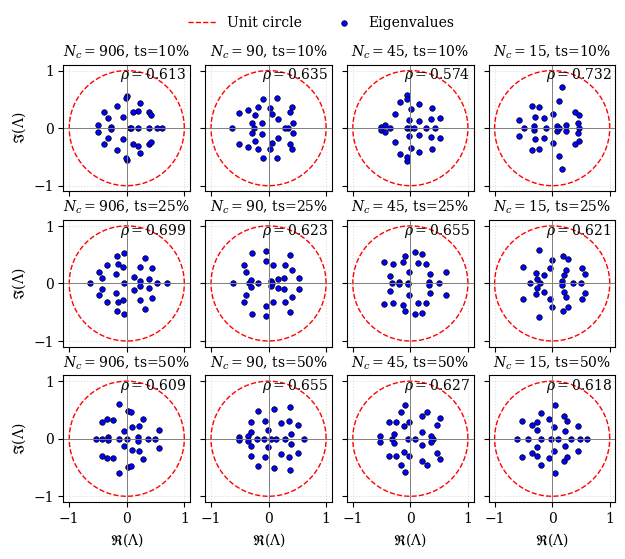}
 \caption{Eigenvalue spectra of the learned Koopman operators for various cycle sampling times ($N_c$) and test set sizes ($ts$). Blue dots denote the eigenvalues in the complex plane, and the red dashed circle is the unit circle ($|\Lambda|=1$).}
 \label{fig:eigA}
\end{figure}

In the decoupled SoC estimation, FNO directly maps $\uin$ with the ground truth maximum discharge capacity per cycle $Q^{\max}$ to the SoC trajectory. Subsequently, the maximum discharge capacity is estimated independently using Koopman for the decoupled version of $Q^{\max}$. A step-wise learning rate scheduler is applied, and to assess the sensitivity of the proposed operator-theoretic framework, we evaluated performance under different resampling granularities. Each raw cycle trace was interpolated to uniform lengths of $N_c\in [906,\, 90,\, 45,\, 15]$ samples per cycle, corresponding respectively to {\it high-resolution, medium-resolution, coarse-resolution, and sub-minute representations}. These different temporal discretizations allow us to test whether the model can maintain accuracy under reduced sampling density, which is vital for practical deployment on systems with limited storage constraints. To gain valuable insight into the stability and generalization characteristics of the learned dynamics models, the eigenvalue distributions relative to the unit circle are observed. Fig.~\ref{fig:eigA} presents the eigenvalue spectra of the learned Koopman operators across $N_c$ and train-test splits (with test set sizes $ts\in [10\%, 25\%, 50\%]$). In each subplot, the eigenvalues (blue dot plots) in the complex plane are shown relative to the unit circle (red dashes). Across all conditions, the eigenvalues lie strictly within the unit circle, confirming the stability and boundedness of the learned dynamics models and the learned modes, a necessary property for reliable long-term predictions. In total, robust, well-generalized models exhibit a compact, well-damped eigen-spectrum, whereas less generalizable models show a dispersed spectrum with near-unstable modes.

% Tables~\ref{tab:singlesoc} and \ref{tab:singleqmax} summarize the prediction accuracies and efficiencies for each model across varying test set sizes and $N_c$. 

\begin{figure*}[!ht]
 \centering
 \includegraphics[width=.85\linewidth]{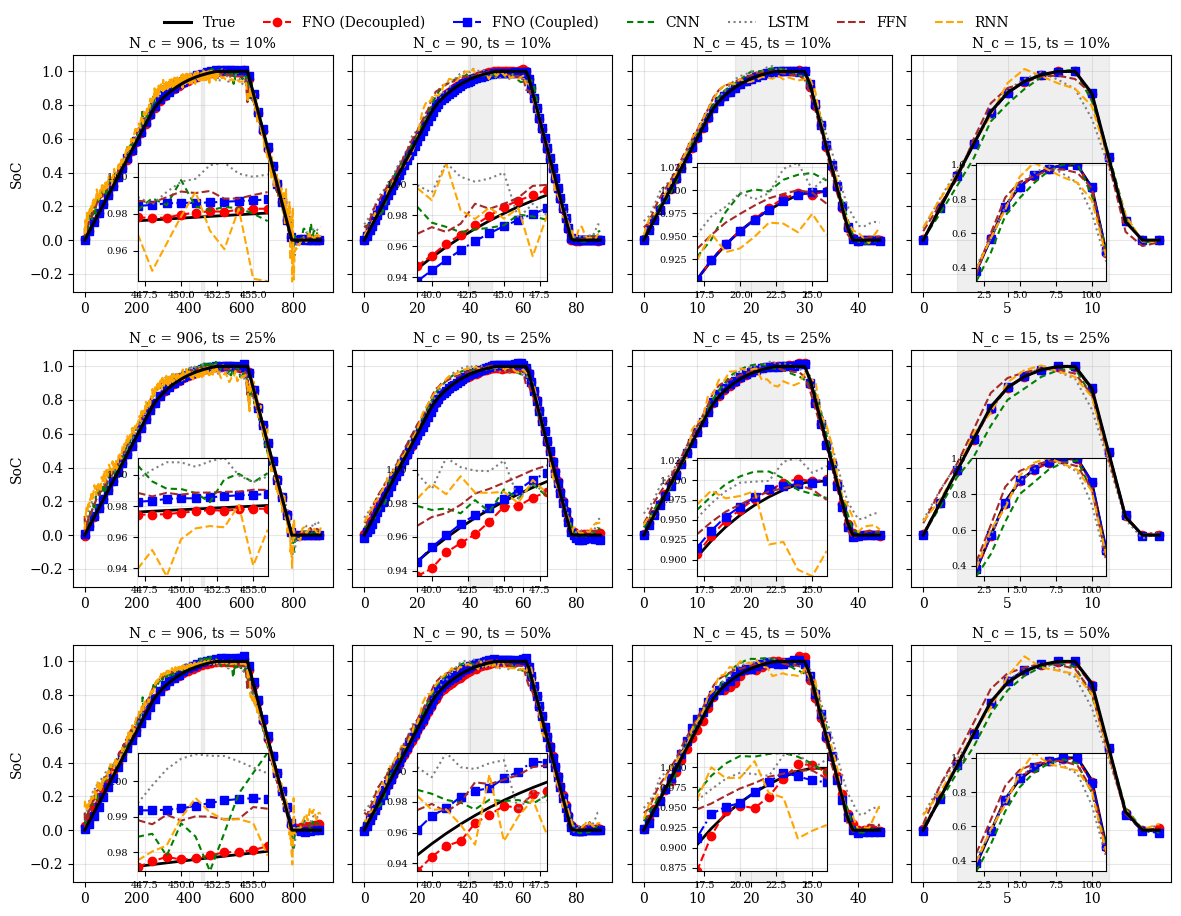} \caption{Plot of one cycle of SoC models (True trajectory (black solid line) versus predictions from the FFN, RNN, BiLSTM, CNN, coupled FNO, and decoupled FNO) across test sizes and cycle counts ($N_c$). A grey-highlighted segment in each plot denotes a zoomed-in region for detailed comparison of the predictions.} \label{fig:socA}
\end{figure*}

Table~\ref{tab:singlesoc} summarizes the SoC prediction accuracies and time efficiencies, while Fig.~\ref{fig:socA} illustrates one cycle of SoC prediction results for each model across varying test set sizes and $N_c$.  

\begin{table*}[!hbt]
\centering
\caption{Comparison of prediction of SoC across test sizes and cycle counts ($N_c$). RMSE, MAE, and inference times are reported. Test cycles are contiguous.}\label{tab:singlesoc}
\renewcommand{\arraystretch}{1.1}
\setlength{\tabcolsep}{3pt}
\resizebox{.85\linewidth}{!}{\begin{tabular}{|c|l|ccc|ccc|ccc|ccc|} \hline

\multirow{2}{*}{Test Size} &\multicolumn{1}{c}{\multirow{2}{*}{Model}} &\multicolumn{3}{|c|}{$N_c=906$}&\multicolumn{3}{|c|}{$N_c=90$}&\multicolumn{3}{|c|}{$N_c=45$}&\multicolumn{3}{|c|}{$N_c=15$}\\ \cline{3-14}

&&RMSE($\%$)& MAE($\%$) &Time(s)&RMSE($\%$)& MAE($\%$) &Time(s) &RMSE($\%$)& MAE($\%$) &Time(s) &RMSE($\%$)& MAE($\%$) &Time(s) \\ \hline

\multirow{6}{*}{10\%} &FFN &$3.7316$ &$3.1038$ & $7.55$ &$3.9097$ &$3.2108$ &$1.64$ &$3.6179$ &$3.1112$ &$1.82$ &$3.2378$ &$2.8097$ &$1.52$\\

&RNN &$5.6511$ &$4.5950$ &$10.84$ &$4.1843$ &$3.3236$ &$2.20$ &$4.0656$ &$3.3717$ &$1.73$ &$5.0687$ &$3.9860$ &$1.49$ \\

&BiLSTM &$4.7855$ &$3.8238$ &$18.13$ &$4.8701$ &$3.9484$ &$1.92$ &$5.8768$ &$4.9536$ &$1.17$ &$4.8382$ &$3.6130$ &$0.69$ \\

&CNN &$4.1603$ &$3.2694$ &$8.92$ &$3.3834$ &$2.7098$ &$1.89$ &$2.3715$ &$1.9461$ &$1.49$ &$3.3080$ &$2.6070$ &$1.36$ \\

&FNO(coupled) &$\mathbf{0.5301}$& $\mathbf{0.4606}$& $\mathbf{0.06}$ &$\mathbf{0.7260}$& $\mathbf{0.6269}$ &$\mathbf{0.07}$ &$\mathbf{0.2760}$ &$\mathbf{0.2240}$ &$\mathbf{0.06}$& $\mathbf{0.3034}$& $\mathbf{0.2393}$& $\mathbf{0.06}$\\ 

&FNO (decoupled) &$\mathbf{0.2476}$ &$\mathbf{0.1671}$ &$\mathbf{0.03}$ &$\mathbf{0.2711}$ &$\mathbf{0.1739}$ &$\mathbf{0.03}$ &$\mathbf{0.2988}$ &$\mathbf{0.2012}$ &$\mathbf{0.03}$ &$\mathbf{0.2284}$ &$\mathbf{0.1738}$ &$\mathbf{0.03}$ \\ \hline

\multirow{6}{*}{25\%} &FFN &$3.4706$ &$2.8699$ &$17.30$ &$3.4892$ &$2.8144$ &$3.12$ &$3.3276$ &$2.6273$ &$2.14$ &$3.5038$ &$2.8657$ &$1.44$ \\ 

&RNN &$5.8203$ &$4.5733$ &$25.07$ &$3.7088$ &$3.0952$ &$3.46$ &$5.6465$ &$4.5838$ &$2.37$ &$3.3615$ &$2.5820$ &$1.78$ \\

&BiLSTM &$4.6360$ &$3.6826$ &$42.82$ &$4.2356$ &$3.4671$ &$4.14$ &$5.3709$ &$4.4259$ &$2.39$ &$4.0181$ &$3.1517$ &$1.22$ \\

&CNN &$3.7592$ &$2.8763$ &$17.23$ &$2.9630$ &$2.3350$ &$2.93$ &$3.1165$ &$2.4953$ &$2.61$ &$3.7549$ &$2.8568$ &$1.57$ \\

&FNO (coupled) &$\mathbf{0.3328}$ &$\mathbf{0.2577}$&	$\mathbf{0.15}$ &$\mathbf{0.8991}$& $\mathbf{0.7141}$&	$\mathbf{0.14}$ &$\mathbf{0.7506}$& $\mathbf{0.4983}$& $\mathbf{0.14}$& $\mathbf{0.3109}$&	$\mathbf{0.2694}$& $\mathbf{0.14}$\\ 

&FNO (decoupled) &$\mathbf{0.2121}$ &$\mathbf{0.1591}$ &$\mathbf{0.07}$ &$\mathbf{0.7812}$ &$\mathbf{0.6050}$ &$\mathbf{0.07}$ &$\mathbf{0.6117}$ &$\mathbf{0.4385}$ &$\mathbf{0.07}$& $\mathbf{0.3324}$ &$\mathbf{0.2488}$ &$\mathbf{0.07}$ \\ \hline

\multirow{6}{*}{50\%}&FFN &$3.1454$ &$2.4271$ &$32.34$ &$3.2605$ &$2.5591$ &$4.03$ &$3.3673$ &$2.5201$ &$3.35$ &$2.9085$ &$2.1998$ &$1.80$ \\

&RNN &$4.5282$ &$3.3571$ &$47.43$ &$3.9999$ &$3.2333$ &$5.68$ &$4.8163$ &$3.9706$ &$3.59$ &$3.6962$ &$2.8909$ &$2.20$ \\

&BiLSTM &$4.4253$ &$3.4590$ &$86.61$ &$4.6231$ &$3.7269$ &$9.14$ &$5.8507$ &$4.6721$ &$4.64$ &$4.0742$ &$3.0895$ &$1.93$ \\

&CNN &$3.6204$ &$2.7860$ &$39.46$ &$3.0514$ &$2.3778$ &$4.61$ &$2.8748$ &$2.3081$ &$3.07$ &$2.5690$ &$1.9807$ &$1.97$ \\

&FNO(coupled) &$\mathbf{0.7303}$& $\mathbf{0.5463}$ &$\mathbf{0.29}$ &$\mathbf{0.4600}$& $\mathbf{0.3226}$&	$\mathbf{0.29}$& $\mathbf{1.1115}$&	$\mathbf{0.7984}$& $\mathbf{0.29}$& $\mathbf{0.4887}$&	$\mathbf{0.3307}$& $\mathbf{0.29}$\\ 

&FNO (decoupled) &$\mathbf{0.6696}$ &$\mathbf{0.3979}$ &$\mathbf{0.14}$ &$\mathbf{0.6177}$ &$\mathbf{0.4187}$ &$\mathbf{0.13}$ &$\mathbf{1.0954}$ &$\mathbf{0.7853}$ &$\mathbf{0.13}$&
$\mathbf{0.3809}$ &$\mathbf{0.2710}$ &$\mathbf{0.14}$ \\
\hline
\end{tabular}}
\end{table*}

\begin{table*}[!htb]
\centering
\caption{Comparison of prediction of $Q^{\max}$ across test sizes and cycle counts ($N_c$). RMSE, MAE, and inference times are reported. Test cycles are contiguous.}
\label{tab:singleqmax}
\renewcommand{\arraystretch}{1.1}
\setlength{\tabcolsep}{3pt}
\resizebox{.85\linewidth}{!}{
\begin{tabular}{|c|l|ccc|ccc|ccc|ccc|} \hline

\multirow{2}{*}{Test Size} & \multicolumn{1}{c}{\multirow{2}{*}{Model}} & \multicolumn{3}{|c|}{$N_c=906$} & \multicolumn{3}{|c|}{$N_c=90$} & \multicolumn{3}{|c|}{$N_c=45$} & \multicolumn{3}{|c|}{$N_c=15$} \\ \cline{3-14}

& & RMSE(Ah) & MAE(Ah) & Time(s) & RMSE(Ah) & MAE(Ah) & Time(s) & RMSE(Ah) & MAE(Ah) & Time(s) & RMSE(Ah) & MAE(Ah) & Time(s) \\ \hline

\multirow{6}{*}{10\%}& RNN& $0.0099$ &$0.0075$ &$1.33$& $0.0039$ &$0.0032$& $1.21$& $0.0023$& $0.0020$& $1.29$& $0.0054$& $0.0046$& $1.23$\\

&BiLSTM &$0.0055$ &$0.0042$ &$0.51$ &$0.0030$ &$0.0026$ &$0.51$ &$0.0021$ &$0.0016$ &$0.50$ &$0.0167$ &$0.0149$ &$0.51$\\

&TCN &$0.0218$ &$0.0188$ &$1.84$ &$0.0234$ &$0.0202	$&$1.79$ &$0.0287	$&$0.0245	$&$1.81$ &$0.0055$ &$0.0044$ &$1.75$\\

& CNN &$0.0034$ &$0.0028$ &$0.50$ &$0.0017$ &$0.0014$ &$0.51$ &$0.0019$ &$0.0015$ &$0.50$ &$0.0132$ &$0.0120$ &$0.50$\\

& Koopman (coupled)& $\mathbf{0.0009}$&$\mathbf{0.0007}$& $\mathbf{0.06}$&$\mathbf{0.0009}$& $\mathbf{0.0007}$&$\mathbf{0.07}$& $\mathbf{0.0015}$&$\mathbf{0.0013}$ &$\mathbf{0.06}$ &$\mathbf{0.0010}$& $\mathbf{0.0007}$&$\mathbf{0.06}$ \\

& Koopman (decoupled) & $\mathbf{0.0011}$&$\mathbf{0.0009}$ &$\mathbf{0.02}$&$\mathbf{0.0009}$& $\mathbf{0.0007}$&$\mathbf{0.02}$&$\mathbf{0.0009}$&$\mathbf{0.0007}$&$\mathbf{0.02}$ &$\mathbf{0.0011}$&$\mathbf{0.0008}$&$\mathbf{0.02}$\\  \hline

\multirow{6}{*}{25\%}& RNN& $0.0133$& $0.0085$& $1.19$& $0.0103$& $0.0080$& $1.20$& $0.0066$& $0.0052$& $1.26$& $0.0112$& $0.0063$& $1.22$\\

&BiLSTM &$0.0077$ &$0.0068$ &$0.52$ &$0.0475$ &$0.0267$ &$0.52$ &$0.0143$ &$0.0127$ &$0.54$ &$0.0142$ &$0.0101$ &$0.52$\\

&TCN &$0.0239$ &$0.0174$ &$1.81$ &$0.0092$ &$0.0073	$&$1.85$ &$0.0455	$&$0.0215	$&$1.8227$ &$0.0107	$&$0.0093 $&$1.8113$\\

&CNN &$0.0071$& $0.0058$& $0.52$& $0.4843$& $0.4602$&  $0.51$& $0.0139$& $0.0106$& $0.52$& $0.0070$& $0.0056$& $0.53$\\

& Koopman(coupled) &$\mathbf{0.0015}$ &$\mathbf{0.0013}$& $\mathbf{0.15}$ & $\mathbf{0.0013}$& $\mathbf{0.0010}$& $\mathbf{0.14}$& $\mathbf{0.0015}$& $\mathbf{0.0012}$& $\mathbf{0.14}$& $\mathbf{0.0014}$& $\mathbf{0.0011}$& $\mathbf{0.14}$\\

&Koopman(decoupled) &$\mathbf{0.0014}$&$\mathbf{0.0012}$&$\mathbf{0.04}$& $\mathbf{0.0014}$& $\mathbf{0.0012}$&$\mathbf{0.05}$& $\mathbf{0.0012}$&$\mathbf{0.0009}$&$\mathbf{0.05}$& $\mathbf{0.0012}$ &$\mathbf{0.0009}$ & $\mathbf{0.04}$\\  \hline
 
\multirow{6}{*}{50\%}& RNN& $0.1031$& $0.0745$& $1.25$& $0.0901$& $0.0674$& $1.29$& $0.1676$& $0.1295$& $1.30$& $0.0153$& $0.0122$& $1.30$\\

&BiLSTM& $0.0101$& $0.0081$& $0.58$& $0.0639$& $0.0624$& $0.58$& $0.0468$& $0.0426$& $0.61$& $0.0097$&$ 0.0074$& $0.61$ \\

& TCN &$0.0147$& $0.0099$& $1.80$& $0.1865$& $0.1579$& $1.85$& $0.0771$& $0.0435$& $1.84$& $0.0640$& $0.0500$& $1.82$\\

& CNN &$0.0527$ &$0.0430$ &$0.53$ &$0.0174$ &$0.0138$ &$0.54$& $0.7835$& $0.6792$& $0.54$& $0.0391$& $0.0331$& $0.58$\\

&Koopman(coupled) &$\mathbf{0.0012}$&$\mathbf{0.0010}$&$\mathbf{0.29}$&$\mathbf{0.0012}$& $\mathbf{0.0010}$&$\mathbf{0.29}$&$\mathbf{0.0011}$& $\mathbf{0.0009}$&	$\mathbf{0.29}$ &$\mathbf{0.0013}$&$\mathbf{0.0010}$&$\mathbf{0.29}$ \\

&Koopman(decoupled) & $\mathbf{0.0013}$& $\mathbf{0.0010}$& $\mathbf{0.09}$& $\mathbf{0.0012}$& $\mathbf{0.0009}$ &$\mathbf{0.09}$& $\mathbf{0.0012}$& $\mathbf{0.0010}$&$\mathbf{0.09}$ &$\mathbf{0.0012}$& $\mathbf{0.0009}$& $\mathbf{0.09}$\\
\hline
\end{tabular}}
\end{table*}

\begin{figure*}[!ht]
 \centering
 \includegraphics[width=.85\linewidth]{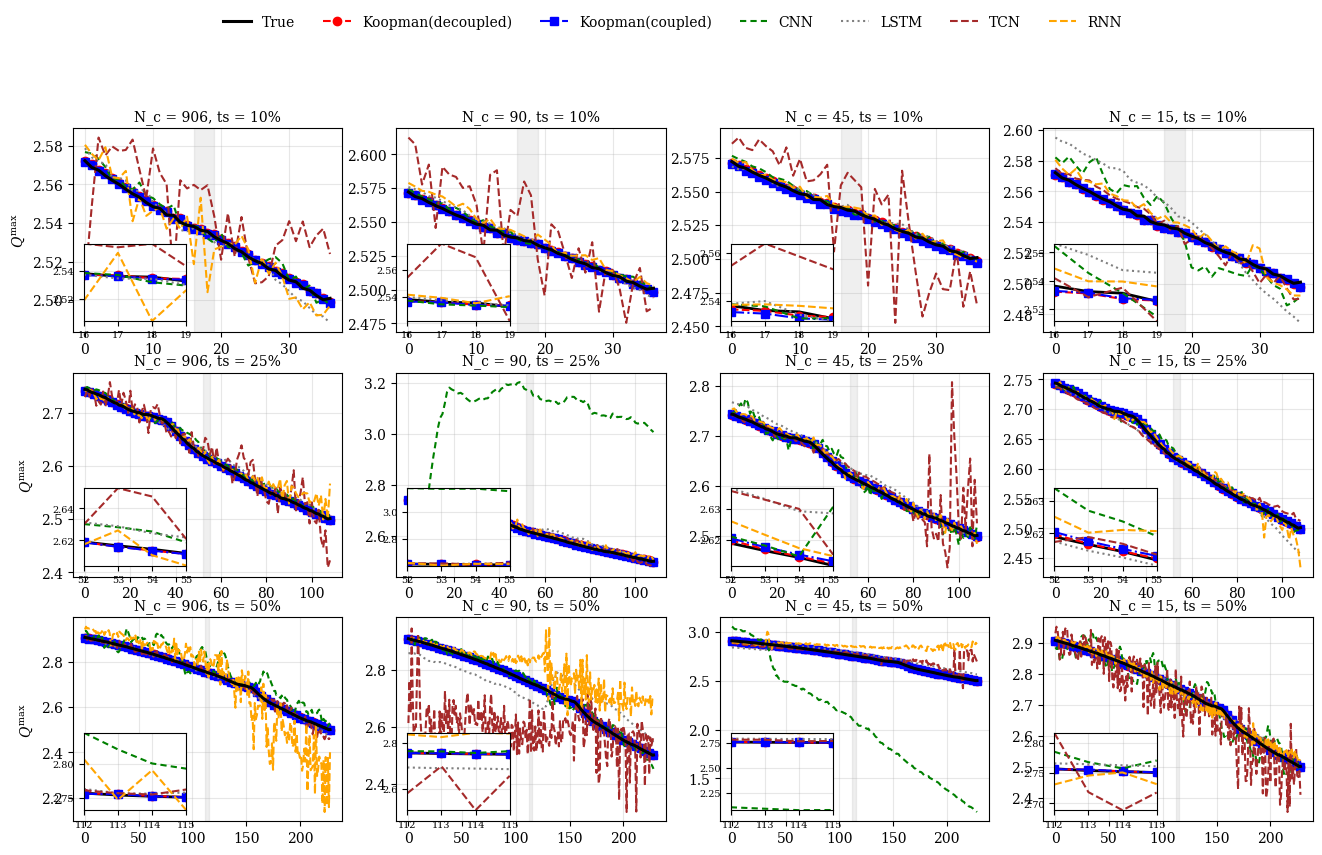}
 \caption{Plot of $Q^{\max}$ (True trajectory (black solid line) versus predictions from the RNN, BiLSTM, TCN, CNN, coupled Koopman, and decoupled Koopman) across test sizes and cycle counts ($N_c$). A grey-highlighted segment in each plot denotes a zoomed-in region for detailed comparison of the predictions. } \label{fig:QA}
\end{figure*}

Overall, FNO models achieve the highest accuracy in SoC estimation under all conditions. Among the conventional models, the CNN and FFN are the next best performers, while the BiLSTM consistently shows the largest error. These accuracy patterns hold as the test size increases: even at a challenging $50\%$ test split, the FNO-based models maintain significantly lower errors than the RNN, LSTM, or CNN. In terms of inference speed, the FNO is faster than the FFN and sequential models, indicating that the FNO models offer an excellent balance of accuracy and efficiency for real-time BMS applications. The decoupled FNO slightly outperforms the coupled FNO in pure SoC error, achieving the lowest overall RMSE in the table. The coupled FNO, meanwhile, delivers nearly comparable SoC accuracy while simultaneously predicting the cell's capacity fade ($Q^{{\max}}$) in one unified model. This integrated approach does not sacrifice much precision; in fact, the coupled FNO still surpasses all the non-FNO benchmarks on SoC metrics, demonstrating that it can capture the interdependence of SoC and state-of-health within a single architecture. In Fig.~\ref{fig:socA}, differences in how all models closely track the actual SoC curve are visible upon zooming in.

Table~\ref{tab:singleqmax} and Fig.~\ref{fig:QA} summarize and illustrate the $Q^{\max}$ prediction accuracies and time efficiencies for each model across varying test set sizes and $N_c$. The proposed Koopman-based estimators, both coupled and decoupled, consistently outperform all baselines (RNN, BiLSTM, TCN, CNN) in terms of accuracy (lowest RMSE and MAE) and inference efficiency (up to an order of magnitude faster). The coupled Koopman model achieves sub-milliamphour RMSE (i.e., $0.0009\; \mathrm{Ah}$) even at $N_c = 15$, reflecting strong data efficiency. Conventional baselines, however, such as BiLSTM and CNN, show reasonable performance under large $N_c$; their accuracy degrades significantly under limited data (i.e., CNN's RMSE increases from $0.0017$ to $0.7835\;\mathrm{Ah}$). In contrast, Koopman models maintain stable performance even under $50\%$ test splits, highlighting strong generalization. These results validate the advantage of operator-theoretic learning in low-data and fast-deployment regimes.

\subsection{Generalization Across Operational Conditions}\label{subsec:generalization}
Having established baseline accuracy and stability, we now evaluate how well the model generalizes under varying operational conditions. Two deployment regimes are evaluated on a held-out battery not seen during pretraining as outlined in \ref{sec:method_gen}:
\textbf{Zero-shot} (no adaptation) \textbf{Few-shot} (k-shot adaptation).  MAE and RMSE are reported for both zero-shot and few-shot, varying $k \in \{1, 5, 10\%\}$. To assess generalization beyond simple random splits, structure-aware partitions aligned with deployment are used:
(i) Temperature Out-of-Distribution (OOD) - hold out one temperature (train: \textbf{B-1, B-2}, test: \textbf{B-3};
(ii) C-rate OOD - hold out one C-rate (train: \textbf{B-4, B-5}, test: \textbf{B-6});
(iii) Chemistry OOD - leave one chemistry cell out (train: \textbf{B-5, B-7}, test: \textbf{B1}). Note that batteries $\textbf{B-1}$ to $\textbf{B-7}$ are reported in Table~\ref{tab:data}.

\begin{table}[ht]
\centering
\renewcommand{\arraystretch}{1.1}
\setlength{\tabcolsep}{3pt}
\caption{OOD generalization comparison in terms of RMSE, MAE, and inference time. Test cycles are contiguous.}\label{tab:ood}
\resizebox{\linewidth}{!}{
\begin{tabular}{|c|c|cc|cc|c|}\hline
\multirow{2}{*}{Scenario} & \multirow{2}{*}{$k$-shot} & \multicolumn{2}{c|}{Soc} & \multicolumn{2}{c|}{$Q^{\max}$}& \multirow{2}{*}{Time(s)} \\ \cline{3-6}
& & RMSE($\%$) & MAE ($\%$) & RMSE (Ah) & MAE (Ah)& \\ \hline\hline

\multirow{4}{*}{Temperature OOD} &$0\%$&$0.8178$&$	0.6215$&$	0.0010$&$	0.0005$&$	1.31$ \\
&$1\%$& $0.6262$&$	0.4395$&$	0.0006$&$	0.0005$&$	1.31$\\ 
&$5\%$& $0.6212$&$	0.4323$&$	0.0005$&$	0.0004$&$	1.27$ \\
& $10\%$& $ 0.6175$&$	0.4479$&$	0.0005$&$	0.0004$&$	1.19$\\ \hline

\multirow{4}{*}{C-rate OOD}&$0\%$& $1.8959$& $1.6150$& $0.0015$&$0.0012$ &$0.24$\\
&$1\%$ &$1.5191$& $1.2454$& $0.0016$& $0.0012$& $0.23$\\
&$5\%$&$0.9885$& $0.7942$& $0.0015$& $0.0012$& $0.23$\\
&$10\%$ &$0.9509$& $	0.7394	$& $0.0015$& $	0.0011	$& $0.20$ \\ \hline

\multirow{4}{*}{Chemistry OOD}&$0\%$&$1.0280	$&$0.7696$&$	0.0011	$&$0.0009$&$	0.55$\\
&$1\%$ &$0.6994$&$0.4871	$&$0.0010$&$	0.0007	$&$0.55$\\
&$5\%$&$0.8066	$&$0.6114$&$	0.0010$&$	0.0009$&$	0.53$\\
&$10\%$ &$0.4803	$&$0.3092$&$	0.0010$&$	0.0006$&$	0.50$\\ \hline

\end{tabular}}
\end{table}
\begin{figure*}
 \centering
 \includegraphics[width=\linewidth]{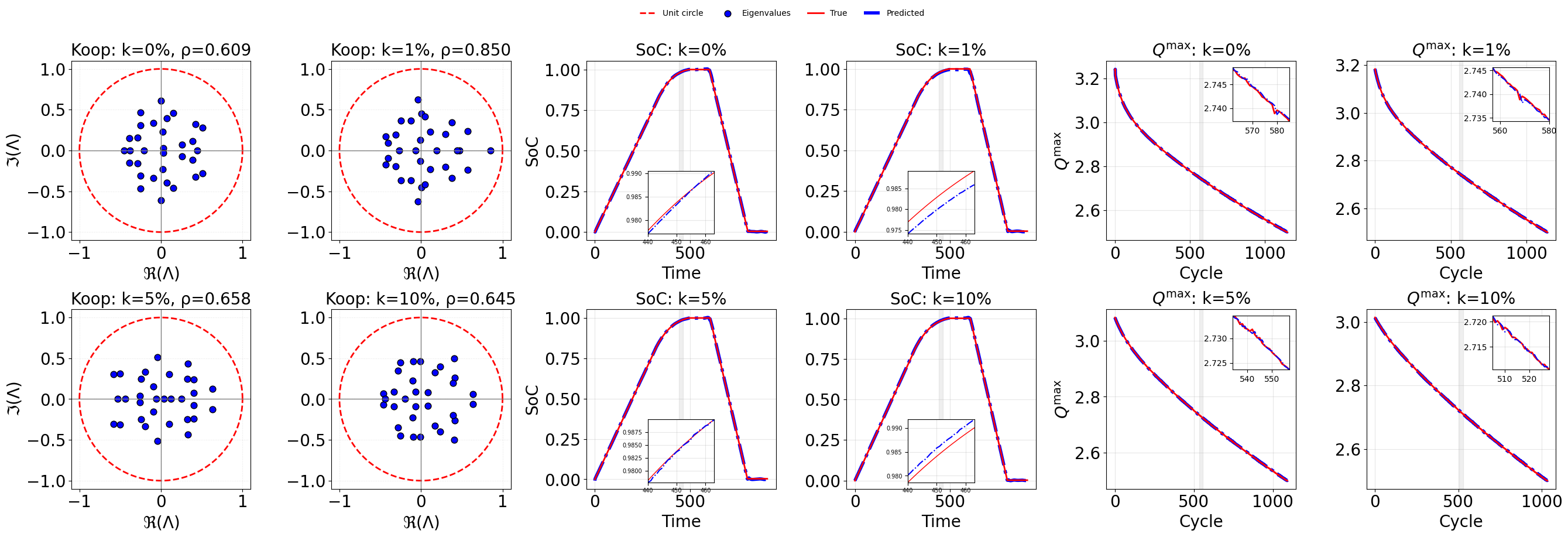}
\caption{\textbf{Temperature OOD Generalization.} \ \textit{Left:} Koopman eigenvalue distributions with increasing $k$ showing spectral radius $\rho$ and stability trends. \ \textit{Middle:} SoC predictions improve with $k$, with insets showing charging accuracy. \ \textit{Right:} Predicted $Q^{\max}$ trajectories with narrowing error as $k$ increases.}
 \label{fig:tempood}
\end{figure*}

\begin{figure*}
 \centering
 \includegraphics[width=\linewidth]{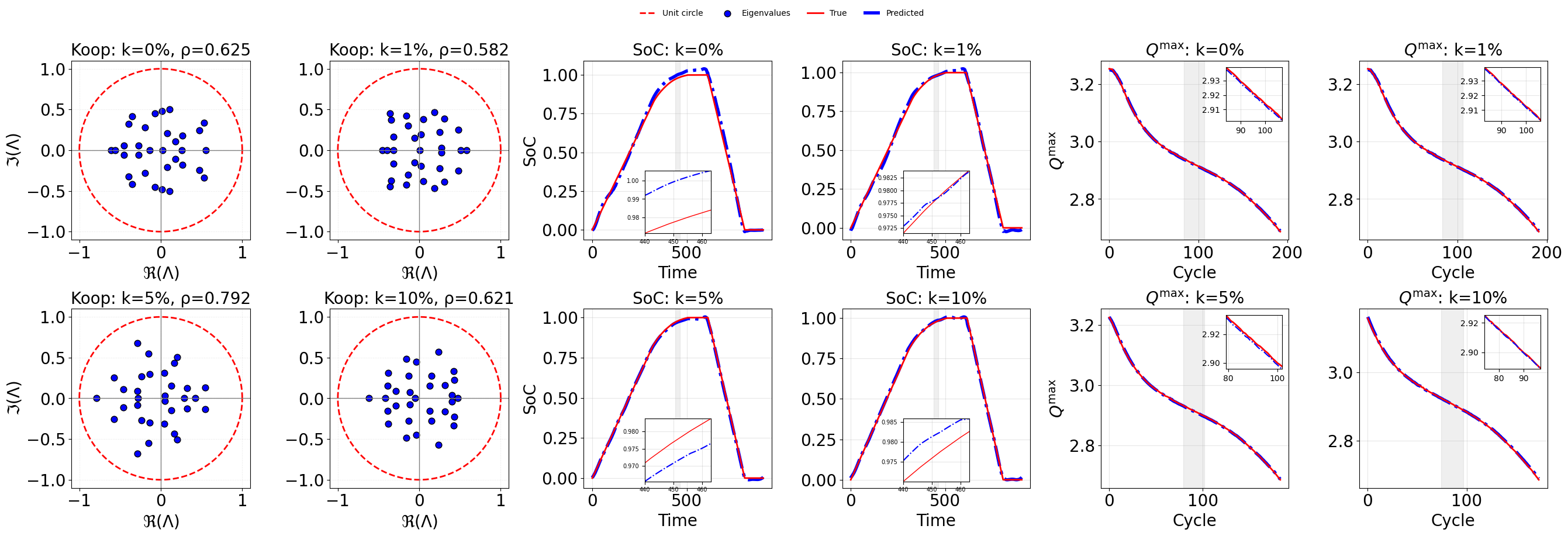}
 \caption{\textbf{C-rate OOD Generalization.} \\textit{Left:} Koopman spectra show improved eigenvalue stability with increasing $k$. \ \textit{Middle:} SoC predictions visibly sharpen with adaptation. \ \textit{Right:} Capacity estimates ($Q^{\max}$) show refinement with adaptation.}
 \label{fig:crateood}
\end{figure*}

\begin{figure*}
 \centering
 \includegraphics[width=\linewidth]{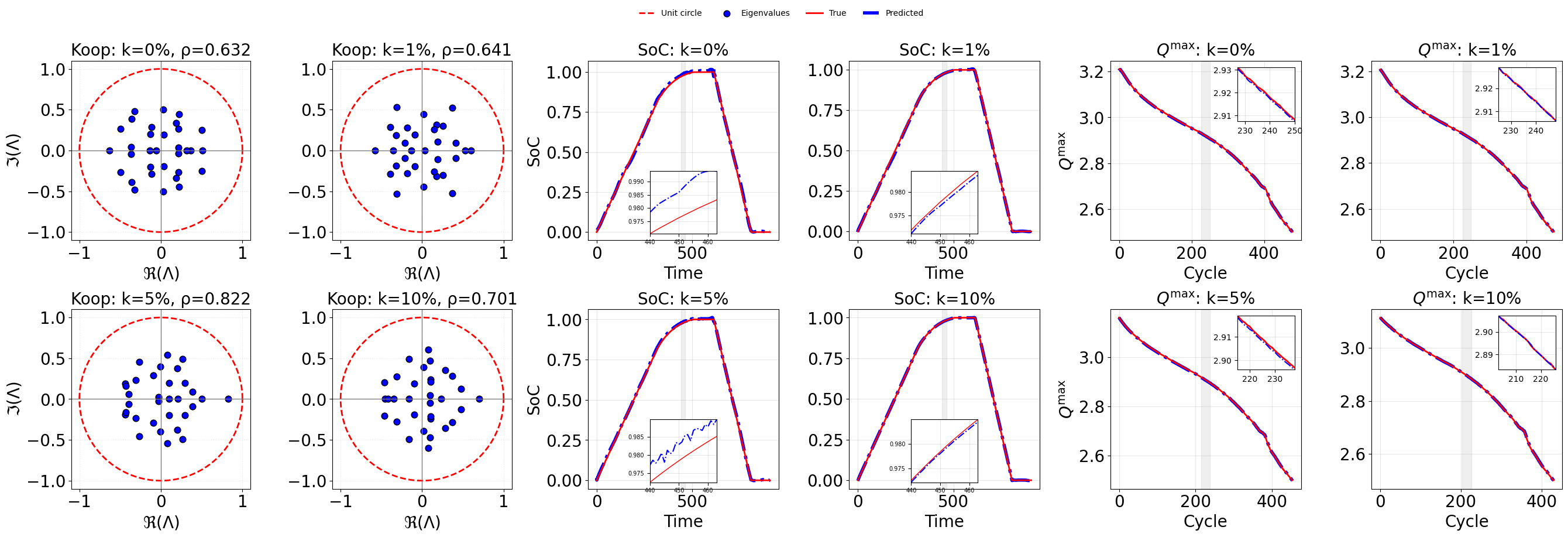}
\caption{\textbf{Chemistry OOD Generalization.} \ \textit{Left:} Koopman eigenvalues maintain structure under unseen chemistries. \ \textit{Middle:} SoC predictions remain robust with late-cycle inset detail. \ \textit{Right:} $Q^{\max}$ predictions consistent across adaptation ratios.}
\label{fig:chemood}
\end{figure*}

The results in Figs.~\ref{fig:tempood}-\ref{fig:chemood} highlight the proposed model's ability to generalize across operational shifts using minimal adaptation. In the temperature OOD setting, both SoC and $Q^{\max}$ predictions remain accurate even under zero-shot conditions, with increasing $k$, and the Koopman operator's spectral radius $\rho$ increases with adaptation, further tightening prediction fidelity and improving latent dynamic encoding. For C-rate OOD, generalization is initially more challenging, as reflected in larger zero-shot SoC errors. However, few-shot adaptation significantly reduces these, particularly during CC-CV transitions. Koopman eigenvalue clusters become more structured and shrink toward the unit circle, reflecting stability. Finally, in chemistry OOD prediction, the model demonstrates compositional transferability. Despite operating under unseen chemistries, SoC and $Q^{\max}$ predictions remain precise. Koopman modes show consistent structure across adaptation levels, validating the framework's robustness to domain shifts. These findings reinforce the utility of latent regularization and explicit operational conditioning in achieving generalization in battery state estimation.

\section{Conclusions}\label{Sec:Conclusions}
This paper presents a joint operator-theoretic framework for estimating battery State of Charge (SoC) and State of Health (SoH), combining a Koopman-inspired capacity predictor with a Fourier neural operator SoC estimator. By injecting maximum capacity trajectories into the SoC pathway and conditioning on control profiles, the architecture overcomes key limitations of prior methods, including dependence on chemistry-specific OCV curves or filter-based heuristics. The framework consistently outperforms classical machine learning models in accuracy and inference speed, while also performing well with generalization across both zero- and few-shot settings, as well as under distribution shifts in temperature, C-rate, and battery chemistry. Importantly, this approach offers a model-agnostic foundation that balances physics-inspired structure with the expressive power of neural operators. By disentangling slow degradation trends from fast SoC fluctuations, it supports robust estimation in scenarios where time scales are decoupled.
Furthermore, the architecture is well-suited for transfer learning because the Koopman operator enables stable extrapolation of long-term capacity evolution, while the FNO provides resolution-invariant SoC estimation that is adaptable to varying sampling rates and domains. This opens promising pathways for extending the framework beyond lithium-ion batteries to broader classes of electrochemical and cyber-physical systems. Future work may explore this transferability or embed more physics-informed information to improve robustness under sparse or degraded measurements further. 

\IEEEpeerreviewmaketitle
\bibliographystyle{IEEEtran}
\bibliography{refb}

\end{document}